\documentclass{emulateapj}
\newcommand{\Jb}{\mbox Jy~beam$^{-1}$}
\newcommand{\muJb}{\mbox{$\mu$Jy~beam$^{-1}$}}
\newcommand{\ex}[1]{\mbox{$\times 10^{#1}$}}

\newcommand{\sn}[1]{\mbox{s$^{#1}$}}

\newcommand{\uv}{\mbox{\em u-v}}
\newcommand{\kms}{\mbox{km s$^{-1}$}}

\newcommand{\HST}{\mbox{\em HST}\/ }  
\newcommand{\Chan}{\mbox{\em Chandra}}

\shorttitle{Crab Wisps in Radio and Optical}
\shortauthors{Bietenholz et al.}

\begin{document}

\title{The Crab Nebula's Wisps in Radio and Optical}

\author{M. F. Bietenholz\altaffilmark{1}, J. J. Hester\altaffilmark{2},
D. A. Frail\altaffilmark{3} and 
N. Bartel\altaffilmark{1}
}

\altaffiltext{1}{Department of Physics and Astronomy, York University,
Toronto, M3J~1P3, Ontario, Canada}
\altaffiltext{2}{Department of Physics and Astronomy, Arizona State
University, Tempe, Arizona, 85287, USA}
\altaffiltext{3}{National Radio Astronomy Observatory, Socorro, New
Mexico, 87801, USA}

\journalinfo{{\em Accepted to the Astrophysical Journal}}
\submitted{{\em On-line materials: } mpeg animations}

\begin{abstract}
We present four new, high-resolution VLA radio images of the Crab
nebula, taken between 2001 February 25 and April 17.  The radio images
show systematic variability in the Crab's radio emission
throughout the region near the pulsar.  The principal geometry of the
variable features is that of elliptical ripples very similar to the
optical wisps.  The radio wisps are seen to move systematically
outward with projected speeds of up to $0.3\,c$.  Comparing the new
radio images to our earlier ones from 1998 and 2000, we show there are
also more slowly moving features somewhat farther away from the
pulsar. In particular, there is a prominent moving feature to the
northwest of the pulsar which has a projected speed of order
$10^4$~\kms.  Striation is seen throughout the nebula, suggesting the
presence of wave-like disturbances propagating through the synchrotron
bubble.  The radio images were taken simultaneously with \HST optical
observations as part of a unique observing campaign to obtain
simultaneous, time-resolved, high-resolution images of the Crab in
different wavebands. Comparing the radio to the optical images, we
find that the radio wisps are sometimes displaced from the optical
ones or have no optical counterparts.  We also find that some optical
wisps in particular, the brightest optical wisps near the pulsar, do
not seem to have radio counterparts.  In the exterior of the nebula,
by contrast, there is generally a good correspondence between the
radio and optical features.
\end{abstract}


\section{Introduction}

The \objectname[]{Crab Nebula} is the most easily accessible pulsar
nebula, and thus an object of singular importance.  Recent results
have highlighted the remarkable nature of the region near the pulsar
in the Crab Nebula.  An image of this region from the \Chan\ X-ray
observatory (Weisskopf et~al.\ 2000) showed a complex geometry
consisting of a tilted toroidal structure with a jet along the
symmetry axis of the torus.  A sequence of images from the Hubble
Space Telescope (\HST) by Hester et al.\ (1996) revealed astonishing
details in this region.  In particular, there is a series of
elliptical ripples in this region, which are usually called ``wisps.''
These wisps have long been known to be variable (Lampland 1921;
Scargle 1969), but only recent observations have shown how dynamic
they really are.  They are variable on time-scales of days, and move
outward rapidly, with speeds up to $0.7\,c$ (Hester et~al.\ 1996;
Tanvir, Thomson, \& Tsikarishvili 1997).  Corresponding rapid motions
are seen in the X-ray (Mori et~al.\ 2002; Hester et al.\ 2002).  In
Bietenholz, Frail, \& Hester (2001; Paper~I hereafter) we showed that
rapid motions in the wisp region were also apparent in the radio, and
that at least one wisp was moving outward with a projected
speed of $0.24c$ over a period of $\sim 2$~months.  Wisps are not
unique to the Crab, since similar radio features have been seen in
Vela (Bietenholz, Frail, \& Hankins 1991), 3C~58 (Frail \& Moffet
1993), and in B1509-58 (Gaensler et al.\ 2001), although none of these
have yet been shown to be as dynamic as those in the Crab.

The exact nature of the wisps is not yet known, but they are generally
thought to be associated with a shock in the wind from the pulsar
which powers the nebula.  As the Crab pulsar spins down, it is
losing $5 \times 10^{38}$~erg of rotational energy per second.  The
bulk of this energy emerges from the pulsar as a highly collimated
wind of relativistic particles and magnetic field.  The particles are
thought to be mostly electrons and positrons, accelerated to
relativistic energies, with possibly a lesser number of ions.  In the
remainder of this paper, we use ``electrons'' to refer to both
electrons and positrons.  This outflow from the pulsar is ultimately
responsible for the synchrotron emission we observe from the body of
the nebula.  The process by which this prodigious transfer of energy
happens, however, is not yet well understood (see e.g., Lyubarsky
2003; Lyutikov 2003;
%C Lozinskaya 2003;  find this one if we're putting it in
Spitkovsky \& Arons, 2004; Arons 2002; Begelman 1999; Hester 1998;
Chedia et~al.\ 1997).
%C we could add Melatos & Melrose, Begelman and others here.

The energy spectrum of the particles in the nebula can be deduced from
the synchrotron emission spectrum, which has a power-law form
extending over many decades of frequency, implying also power-law
energy spectra of the electrons. The mechanism by which this particle
spectrum is generated is still not known.  In particular, most of the
present theories of the particle acceleration process fail to produce
a substantial number of electrons with relatively low energies, i.e.,
with Lorentz factors, $\gamma$, less than $10^4$, which are required
to produce the nebula's radio emission (e.g., Arons 2002; Atoyan
1999). The electrons with $\gamma < 10^4$ dominate by number, although
the electrons with $\gamma > 10^4$, which produce the nebular optical
and X-ray emission dominate the energy of the population.  The current
radio brightness of the Crab requires a time-averaged electron
injection rate over the lifetime of the nebula of $10^{40.5}$~\sn{-1}.
%C number from Lyubarsky, 2003 (10^40to41) or Gallant et al Boston conf

Studying the wisps is of particular importance because they are the
only observational window on the neutron-star nebular interface.
The wisps are now generally thought to be associated with the shock
which terminates the highly collimated outflow from the pulsar and
randomizes it before it enters the body of the nebula.  It is also
in this region that the nebular particle spectrum is established.
Multi-wavelength observations may therefore provide observational
constraints which can help answer the basic questions of how the power
is transferred from the pulsar into the nebula and how the nebular
particle spectrum is established.  A synchronized multi-wavelength
campaign of time-resolved observations of the Crab was undertaken
using NRAO's VLA\footnote{The NRAO Very Large Array is a facility of
the National Science Foundation operated under cooperative agreement
by Associated Universities, Inc.}, the \HST\ and the \Chan\ X-ray
satellite.

The radio observations are of special interest because, as mentioned,
the origin of the radio emitting electrons presents a particular
problem.  The continuity of the emission spectrum from the radio to
the optical suggests that the radio-emitting electrons are produced by
the same mechanism as the higher energy ones, and the uniformity of
the radio spectral index over the nebula suggests that only a single
acceleration mechanism produces all or most of the radio-emitting
electrons (Bietenholz et al.\ 1997).  Understanding the acceleration
process, therefore, will require understanding the origin of the
majority  population of $\gamma<10^4$ radio-emitting electrons.
%Cradio-emitting > 10^4

In this paper, we will discuss chiefly the radio results and the
comparison between the radio and optical results of the
multi-wavelength campaign mentioned above.  A discussion of the
optical and X-ray observations can be found in Hester et~al.\ 2002 and
Mori et~al.\ 2002.  We describe our observations in \S\ref{sdatredux}.
We show a selection of representative images and describe the
time-evolution of the radio synchrotron emission in \S\ref{sresults}.
In \S\ref{srocomp}, we compare the radio images and their
time-evolution to the optical and X-ray results, and finally in
\S\ref{sdiscuss}, we discuss our findings.

\section{Observations and Data Reduction}
\label{sdatredux}

The radio observations of the Crab Nebula were carried out with the
VLA in the 5~GHz band.  We observed approximately every two weeks
between 2001 February and 2001 April.  Table~\ref{tobs} shows details
of our four observing runs, all of which were done using the VLA in
the B configuration. In all four sessions, we observed at frequencies
of 4615 and 4885~MHz with a bandwidth of 25~MHz, used 0556+238 as a
phase calibrator, and calibrated the amplitude scale using
observations of both 3C~48 and 3C~286.  The total time per observing
session was $\sim$12~hours.  We chose to observe at a spaced pair of
frequencies within the 5~GHz band for better \uv~coverage, which is
critical when imaging an object as extended as the Crab.

Our goal was to obtain time-resolved observations, which would
complement those in the optical with the \HST\ and in the
X-ray with the \Chan\ X-ray satellite.  Time resolved radio
observations with the VLA present the problem that the Crab Nebula
exhibits structure at spatial scales from $\sim$6\arcmin\ down to
$<$1\arcsec.  This range of spatial scales is not well sampled by any
single VLA array configuration, and the B array configuration we used
samples only spatial scales between $\sim$50\arcsec\ and
$\sim$1\farcs4\ at 5~GHz.
%C  In Paper I we HP filter with 20\arcsec - not sure why we are
%C  filtering w/ 14\arcsec\ here - o well
The usual approach of recovering the large scale structure by
additional observations using more compact array configurations is
precluded by the fact that VLA configuration changes occur only once
every four months.

We devised a strategy for obtaining reliable observations using only a
single array configuration in Paper~I, and we repeat a brief
description here.  Any rapid time evolution must occur on the smaller
spatial scales, since a speed of $c$\/ represents a proper motion of
only $\sim$3\arcsec\ per month.  Our solution to the problem of the
large scale structure is to use maximum entropy deconvolution (AIPS
task VTESS; see Cornwell \& Evans 1985, Cornwell 1988) by supplying a
low resolution support (default) in the deconvolution process.  The
deconvolved image is then biased to be as close to the support as is
allowed by the data.  We chose this support to be the same for all
four of our epochs, and we used the same default image that was used
in Paper~I (made from data taken in the B,C and D array configurations in 1980-81;
Bietenholz \& Kronberg, 1990, 1991).  We did scale the support
according to the known radio expansion of 0.13\%~year$^{-1}$
(Bietenholz et al., 1991), although the effect is small over our
observing period.  The use of the same default for each session serves
to make any differences between our images at different epochs be only
those that are demanded by the data.

We also compare the images from 2001 to our earlier images from
Paper~I, taken on 1998 August 10, October 13, and 2000 February 11, in
order to determine whether there are any longer timescale variations.
The reduction of these earlier data was essentially the same as for
the present data\footnote{We note that for the 2000 February
11 epoch, we used the image from 1998 October 13 rather than our usual
default image.  The reason for this was that this data set had
slightly different \uv~coverage, and the deconvolution errors appeared
to be larger. Using the 1998 October 13 image as a default reduced the
spurious differences from the other images.  These spurious
differences are relatively easy to distinguish from the true changes,
so our results are not dependent on the exact choice of the default
image.}, and is more fully described in Paper~I\@.  We re-imaged the data
using the same grid spacing as was used for the 2001 data.

Our process of deconvolving with a default image causes the structure
in our radio images on the large spatial scales, i.e.\ those
$>$50\arcsec, to be derived predominately from the support image, and
thus from considerably earlier multi-configuration data.  This
larger-scale structure is thus not directly comparable to that in the
optical and X-ray observations.  However, as we argued above and in
Paper~I, there is little reason to expect rapid variations on these
%C dale comments: c 17months = 50\"
scales.  Our chief interest in this paper is in the smaller scale
structure.  On spatial scales smaller than $\sim$20\arcsec\ our
\uv~coverage is excellent, and we believe that our images reliably
indicate the structure of the nebula, and in particular, changes
therein from one epoch to another.  To further ensure the reliability
of the images we present, we work for the remainder of this paper only
with spatially high-pass filtered radio images.  We conservatively
filtered with a Gaussian of full width at half-maximum (FWHM)
14\arcsec, which serves to isolate those spatial features for which we
have good \uv~coverage and thus reliable imaging. The high-pass
filtering will ensure that our images at each epoch contain only
structure reliably determined by the data and not derived from the
support, and will reduce any spurious differences between the images
due to possible deconvolution errors.

The optical observations were carried out with the \HST\/ Wide Field
Planetary Camera~2 (WFPC2) using a relatively line-free filter
centered near 5500~\AA\ (F547M).  They consisted of 24 visits between
2000 August and 2001 April at 11 day intervals.  The radio
observations occurred during the period between \HST\ visits 19 to 24.
%C taken from Hester's 2002 ApJL paper
The X-ray observations were carried out using the \Chan\ Advanced CCD
Imaging Spectrometer (ACIS).  The optical and X-ray observations are
more fully described in Mori et al.\ 2002 and Hester et~al.\ 2002.
%C of xxx
%C 
%C  Optical: the ``u50v...'' is the original filename of the
%C  images I have
%C 10feb2001 = u50v1801r_cvt.c0h
%C 21feb2001 = u50v1901r_cvt.c0h
%C 17apr2001 = u50v2401r_cvt.c0h
%C - conclude these are visits 18,19 and 24
%C  The sprite is quite prominent (both X-ray and optical)
%C  in visit H19-23, ie over our whole observing period
%C 
%C xxx Todo Poloidal loops HST paper
%C 
%C  radio   getn 157, pixr -0.005,0.006
%C  optical getn 132; pixr 60,120
%C Sprite:
%C  layers r=4,o=1 -> seems to be some sprite stuff
%C  layers r=5,o=3 -> seems to be some sprite stuff
%C  layers r=7,o=6 -> seems to be some sprite stuff
%C 

\section{Results}
\label{sresults}

\begin{figure}
\plotone{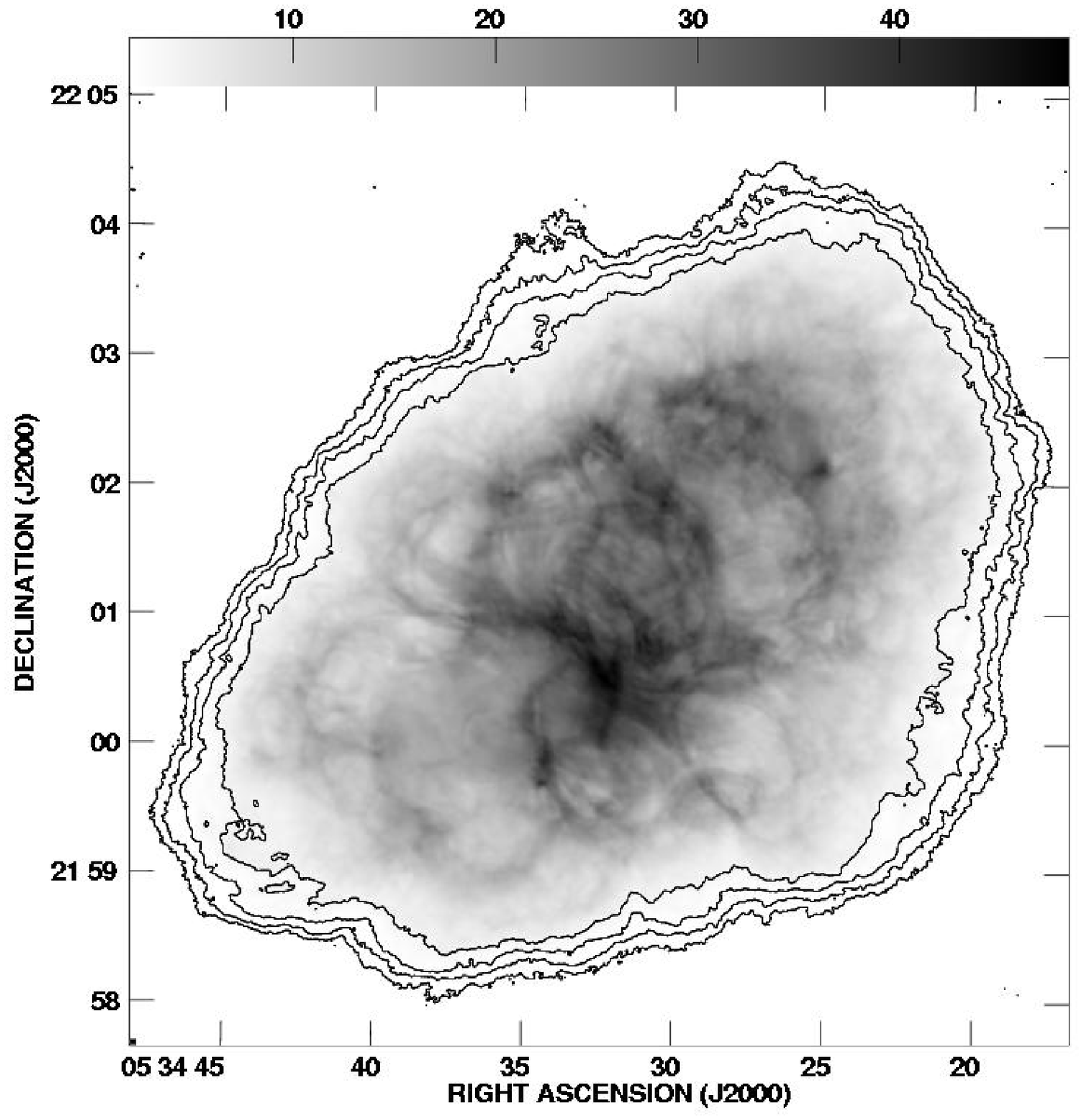} 
% corresponds to CRBAP01 AS3 .VTCPB   .22
% use lpen=2 in LWPLA
\caption{The Crab at 5~GHz on 2001 April 17.  The FWHM size of the
restoring beam was 1\farcs4.  The peak brightness is 48.5~m\Jb, and
the image has been corrected for the primary beam response of the VLA
antennas.  The rms background was 48~\muJb\/ before primary beam
correction.
%C TVSTAT rms near exterior is about 70 muJb after PBCOR
The contours are drawn at 0.75, 2, 4 and 8\% of the peak, and the
greyscale is labelled in m\Jb.  Maximum entropy deconvolution was used
with a support to recover the low spatial frequency structure (see
text, \S~\ref{sdatredux}) for details.
%C  corresponds to crab-radio-all.mpg 
% *** This corresponds to video1.mpg ***
The accompanying animation shows a 61\arcsec\ $\times$ 70\arcsec\
region near the center of the nebula between 1998 August 10 and 2001
April 17 (see Table~\ref{tobs}).  The mpeg time-scale is non-linear,
being compressed between 1998 and 2000.  The individual frames have
been high-pass filtered with a 14\arcsec\ FWHM Gaussian filter to
enhance the mobile features.
\label{faprimg}}
\end{figure}

In Figure~\ref{faprimg}, we show our latest radio image of the Crab
Nebula, taken on 16 April 2001.  The size of the restoring beam was
1\farcs4 FWHM, which was conservatively chosen as a common size for
all seven epochs.  The peak brightness was 49~m\Jb, and the rms
background level was 48~\muJb\ before applying the primary beam
%C rms refers to CRBAP01 ??; TVSTAT of a region near
%C suggests an exterior rms of about 70 uJy after PBCOR
correction, which causes the rms background to rise with the
distance from the center of the image, to $\sim$65~\muJb\ near the
edge of the nebula.  Table~\ref{tobs} gives the peak brightness and
rms background for all our images.

%C Table has been updated for 2001 obs; midpoint dates
%\begin{deluxetable}{r@{ }l@{ }l c c c c}
\begin{deluxetable*}{r@{ }l@{ }l c c c c}
%C PP \tablewidth{0pt}
\tablecaption{Observing Run and Image Details\label{tobs}}
%  wierd, this label seems to need to be here to work
\tablehead{
\multicolumn{3}{c}{Date (midpoint)} & \colhead{Length} & \colhead{Midpoint}  &
\colhead{Image Peak Brightness} & \colhead{rms background\tablenotemark{a}} \\
%C \colhead{} & \colhead{}  & \colhead{} & \colhead{}
  & & &
\colhead{(hrs)}  & \colhead{(days since 1998 Aug 10)} &
\colhead{(m\Jb)} & \colhead{(\muJb)} 
}
%C            date    midp   day 1998.0   since 1st  
%C  day no: 1998 Aug  9.65 ->  221.65         0
%C          1998 Oct 13.46 ->  286.46     64.81
%C          2000 Feb 12.10 ->  773.10    551.45
%C          2001 Feb 25.10 -> 1152.10    930.45
%C          2001 Mar 13.06 -> 1168.06    946.41
%C          2001 Mar 27.02 -> 1182.02    960.37
%C          2001 Apr 16.95 -> 1202.95    981.30
%C  1-7 = 981.30 days = 2.69 avg. years
%C  4-7 =  50.85 days 
\startdata
%C  put in time-reverse order so we can fill in the earlier ones
2001 & Apr & 17    & 12.1     &  981 & 4.83 & 48 \\
2001 & Mar & 27    & 12.5     &  960 & 4.83 & 73 \\
2001 & Mar & 13    & 12.5     &  946 & 4.81 & 93 \\
2001 & Feb & 25    & 12.5     &  930 & 4.80 & 77 \\
\multicolumn{6}{l}{\dots Additional epochs from Paper I} \\
%C  check that the dates/day numbers all add up; the ones
2000 & Feb & 11    & \phn 9.4 &  551 & 4.77 & 76 \\
1998 & Oct & 13    & 10.1  & \phn 65 & 4.81 & 82 \\
1998 & Aug & 10 & \phn 9.6 & \phn\phn 0 & 4.70 & 100 \\
\enddata
\tablenotetext{a}{The listed values are before the VLA primary beam
correction.  Making this correction will cause the rms background
level to rise farther away from the pulsar, so that the true values
near the edge of the nebula are $\sim$35\% higher than those listed.}
\end{deluxetable*}

Our images at the other epochs in 2001 are very similar to the one
shown in Figure~\ref{faprimg}, and we do not reproduce them here (an
image from 2000 can be seen in Paper~I).  To clearly show the
variation, we include an animation accompanying Figure~\ref{faprimg},
showing the variation in the central region over all seven epochs of
radio observations.  In order to discuss the variable features in the
nebula which are the focus of this paper, we proceed to make {\em
difference}\/ images by subtracting the image at an earlier epoch from
that at a later one.

\subsection{Variability in the Radio Images\label{sradvar}}

\begin{figure}
\plotone{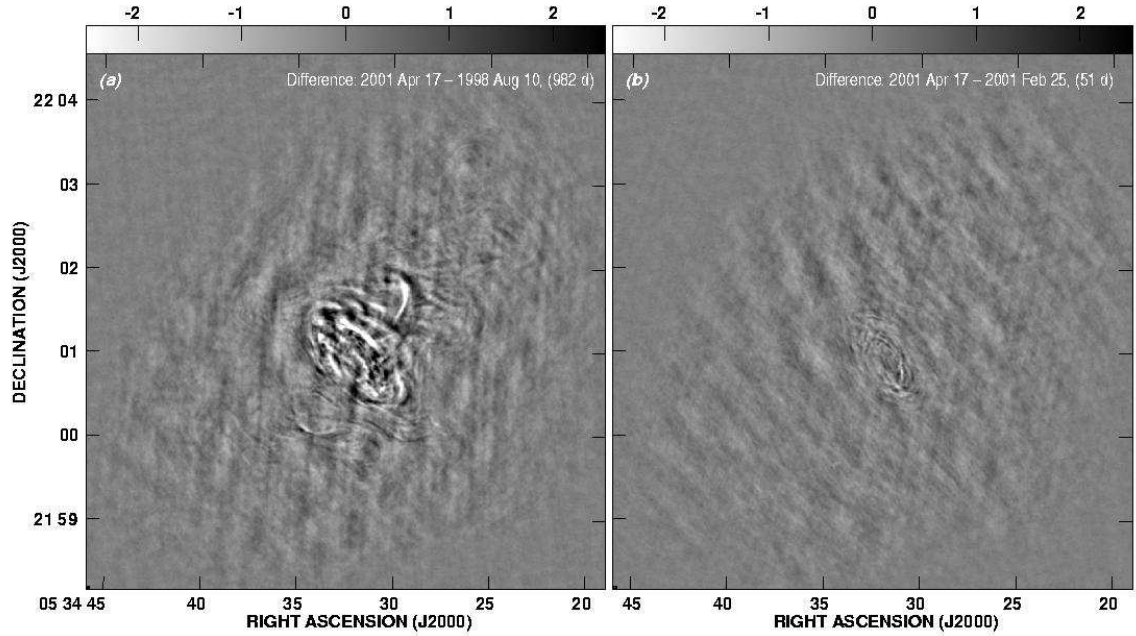}
% made from CRB-ALLM L17.LYRDFC.   1
% made from CRB-ALLM L47.LYRDFC.   1
% blc 101 101; trc 1350,1380;  pixr -2.5 to +2.5 mJy (same for both)
%C These have the better version of Feb2000; just use linear scaling.
\caption{Difference images showing the changes occurring in the
radio nebula, both on time-scales of a few years and about two months.
Moving ripples are apparent throughout the central region.  The
greyscale is labelled in m\Jb\ and is the same in both panels.  The
individual images were high-pass filtered with a 14\arcsec\ FWHM
Gaussian filter before forming the differences (see text
\S~\ref{sdatredux}).  The FWHM of the restoring beam is 1\farcs4.  The
compact, well-delimited features near the center of the nebula are
real, while the undulations on larger scales of $>14$\arcsec\ are likely
artifacts due to deconvolution errors (see text, \S~\ref{suncert}).
%C  The 1, 5 and 10\% contours of the total intensity image of 2001
%C February 25 are included for reference. 
($a$): the 2001 April 17 image minus the 1998 August 10 image, i.e.,
a difference image over an interval of $\sim$3~year or 982~days.  
%C The 1998 August 10 image was scaled by up 1.0035 to account for the
%C general expansion of the Nebula before forming the difference.
%
($b$): the 2001 April 17 $-$ 2001 Feb 25 difference image, i.e., over
an interval of $\sim$2~months or 51~days.
\label{fdifimg}} 
\end{figure}

We first discuss the longer term variations in the radio images.
The image in Figure~\ref{fdifimg}$a$ shows the difference between 2001
April 17 and 1998 August 10, showing the changes over an interval of
$\sim$3~years (981~days)\footnote{ We expanded the 1998 image by a
factor of 1.0035 to account for the overall expansion of the radio
nebula of 0.13\% per year (Bietenholz et al.\ 1991) before forming the
difference image.}.  Figure~\ref{fdifimg}$b$ shows the difference
between the 2001 April 10 and 2001 February 25 images, showing the
changes over a shorter interval of $\sim$2 months (51~days).
%C we neglect the overall
%C expansion of the nebula over this short interval).  
The most prominent differences over both time periods are in the
central region of the nebula and consist principally of ripples
with an approximately elliptical geometry.  These ripples are very
similar to what we reported in Paper~I\@ where we first showed the
elliptical geometry of the radio variation over intervals of $\sim
2$~months and $\sim 1.5$~years.

With the present data we have better time-sampling over both longer
and shorter time-scales, and can more reliably determine the motions
over both long and short time-scales.  The differences over the longer
time interval of $\sim$3~years are considerably larger than those over
the shorter interval of $\sim$2~months, especially near the center of
the nebula.  The average rms variability of the central region ($2\arcmin
\times 2\arcmin$ centered on the pulsar) over the $\sim$3~year
interval is 0.4~m\Jb\, while that over the $\sim$2~month period is
only 0.2~m\Jb, representing a fractional rms variability of 1.3\% and
0.7\%, respectively.  The observation of larger amplitude differences
over longer time scales is not dependent on the particular selection
of images: the differences between 2001 March 27 and 1998 October 13
are similar in amplitude to those in Figure~\ref{fdifimg}$a$, while
those between 1998 October 13 and 1998 August 10 are similar to those in
Figure~\ref{fdifimg}$a$.
%C  here are the stats on the diff. images from CRAB-ALLM .CUBHP .1 
%C epoch/layer     extr      rms_big  rms_cen
%C               --------all in mJy/bm--------
%C  7-1 (a)      -5.2 +4.1    0.458   1.15  (unscaled for expansion)
%C  7-1 (a)      -5.2 +4.1    0.430   1.14  (scaled for expansion)
%C  6-2          -5.2 +4.0    0.402   1.09
%C  7-4 (b)      -2.0 +1.8    0.220   0.350
%C  2-1          -3.5 +2.4    0.229   0.476
%C rms_big = blc 241,238;  trc 1252,1198, 972532 pix, 39405 bareas
%C rms_cen = blc 619 529;  trc  883, 787,  68635 pix,  2780 bareas
%C  If I expand layer 1 by 1.0035 (which is 0.13%/yr * 981d/365.25d)
%C  the rms (expanded-normal; same epoch!) is  
%C               -1.8 +1.0    0.141   0.078 
%C  however, a better comparison is some exterior regions
%C 
%C If we assume that the rms variation outside
%C the central $2\arcmin \times 2\arcmin$ is representative of the
%C variation due to deconvolution errors and noise, we can estimate the
%C true rms time variability in the central region to be 0.3~m\Jb\ over
%C all seven images and 0.1~m\Jb\ over 2001 only.  This is 1.0 and 0.3\%
%C respectively of the average surface brightness over this region.  
A significant component of the variability has a
timescale longer than the two month observing period in 2001.  In
particular, this suggests significant variability on a time scale of
a few years, long compared to the rapid wisp motions, but short
compared to the age of the Nebula. 

The geometry of the variability in the central region consists
principally of a series of elliptical arcs, remarkably similar to the
geometry of the elliptical wisps seen in the optical and  X-ray
(Hester et al.\ 1995, 1996, 2002; Mori et~al.\ 2002; Weisskopf
et~al.\ 2000).
%C The most prominently variable features are to the northwest of the
%C pulsar, the same side on which the brightest optical wisps are
%C located.  
In general, the difference features seem to consist of positive ridges
lying farther from the pulsar than parallel negative ones, implying
outward motion.  The animations accompanying Figure~\ref{faprimg} show
more clearly that the motion is generally directly outward.
%C  a little better now but that's still kinda weak still

On the difference image over the $\sim$3~year interval
(Fig.~\ref{fdifimg}$a$), there are additional striations visible over
most of the body of the nebula. The geometry is still generally
arc-like but no longer centered on the pulsar.  The pattern of
positive ridges lying farther from the pulsar than negative ones is
less readily discernible in this region, but the most prominent ridges
still show this pattern, again implying generally outward motion,
albeit at lower velocities and amplitudes than near the pulsar.
%C  not sure this is even true

Also on this $\sim$3~year difference image, there is a prominent
feature which does not share the elliptical geometry of the wisps.  It
is somewhat beyond the wisp region, to the northwest (NW) of the
pulsar, and is in roughly the same orientation but on the opposite
side of the jet seen in the X-ray and optical.
%C is jet seen or just inferred in optical
We will call this feature the ``moving arc''.  It extends from about
44\arcsec\ to 68\arcsec, or 1.3 to 2.0\ex{18}~cm, from the
pulsar\footnote{We adopt a distance of 2.0~kpc throughout this paper,
(Trimble \& Woltjer 1971)}.
%C Distance to Crab:
%C  Trimble & Woltjer 1971: 2 +/- 0.25 kpc from proper mot/radial mot
%C        of filaments
%C  Manchester & Taylor, Frail & Weisberg 1990 and Green's
%C   catalog all cite them
%C
%C length of counter-jet: 44.2, 68.3" just done off the total rms image
%
The part nearer the pulsar seems to be moving transversely, while the
more distant part is moving away from the pulsar.  The proper motions
are difficult to estimate reliably because of the difficulty in
isolating the feature in question on the images.  From the difference
images,
%C   this is what we got off the non-difference images
%C ----------
%C we estimate a displacement of $\sim$1\arcsec\ transversely for the end
%C nearer the pulsar, and $\sim$2\arcsec\ radially for the end farther
%C from the pulsar, corresponding to projected speeds of 3500 and
%C 7000~\kms\ respectively.  
%C --------
we estimate projected speeds of $\sim$7500~\kms\ transversely for the
end nearer to the pulsar and $\sim$14,000~\kms\ radially for the end
farther from the pulsar.  These speeds should be regarded as somewhat
uncertain, since there are probably changes in brightness that are not
associated with motion in the moving arc.

\subsection{Uncertainties in the Difference 
Images\label{suncert}}

Are these features in the difference images real?  Deconvolution
errors, caused by un-sampled parts of the \uv~plane are often larger
than the image rms background (Briggs, Schwab, \& Sramek 1999).  The
deconvolution errors are driven by the noise, so they may be different
from one epoch to another even in the case of identical \uv~coverage.
They tend, however, to be confined to poorly sampled areas in the
\uv~plane.  The result in the image plane is undulations which cover
much of the deconvolved area.  Compact, well-delimited features such
as those observed near the center of the Crab Nebula are not likely to
be due to deconvolution errors, and neither are the arc-like
striations seen outside the central region.  There are some
undulations on a larger scale in both difference images, however,
which likely are the result of deconvolution errors.

\section{Comparison to the Simultaneous HST images}
\label{srocomp}

We now turn to a comparison of the radio images of 2001 to the
(almost) simultaneous \HST\ optical ones. For such a comparison, we
must first accurately register the radio and optical images, which is
most easily done using the pulsar.  The pulsar is identifiable on some
of the radio images, particularly that of 2001 March 27.  The average
radio flux density of the pulsar is $\sim 0.5$~mJy and is quite
variable (Moffett \& Hankins 1996),
%C  I used to have 1.5 mJy from Lorimer but that doesn't make sens
%C  14.4 mJy +/- 3.2 @ 1.4 GHz, 45+/10 @ 925MHz, 686+/-70@408MHz
%C    I get 0.5 +/- 0.1 from these numbers.
consistent with what is visible in our images (we note that the pulsar
is clearly visible when an image of the rms variation over the radio
images is formed as we do below in Figure~\ref{fro-rms}$a$). We
identify the brightness peak of the pulsar in the radio images with
that in the \HST\ images.  By using the pulsar we are ensured of an
accurate registration, not dependent on the absolute astrometry of
either the radio or the \HST\ images.  The radio astrometric scale and rotation
are such that the relative positional accuracy should be a few milliarcsec.  The
\HST\ astrometric scale and rotation are also well known, with estimated relative
positional uncertainties over the entire image of $<$0\farcs1\
(Baggett et al.\ 2002).  We therefore estimate that the radio-optical
alignment is accurate to 0\farcs1\ or better.
%C  Richards, E. A, Kellerman, K. I., Fomalont, E. B., Windhorst,
%C  R. A., \& Partridge, R. B. 1998, \aj, 116, 1039
%C  These guys say that HST pixel scale and rotation are well
%C   known and don't worry about it over small (HST?) fields
%C  Fine guidance system of HST accurate to a few mas, but HST
%C   guide stars only accurate to 1 to 2\"

%
%C \footnote{A check of the positions of stars 18, 19, 21 and 24
%C from Wykoff \& Murray (1977), which can be seen on the \HST\ images,
%C shows that the \HST\ scaling and rotation consistent with that of
%C Wykoff \& Murray to 0.5\% and 0.3\arcdeg\ respectively}.  Since no
%C stars are visible on the radio images, we must rely on the scaling and
%C rotation on the VLA images which is believed to be accurate to 
%C  footnote 7: CCD delta I delta dec

\begin{figure}
\plotone{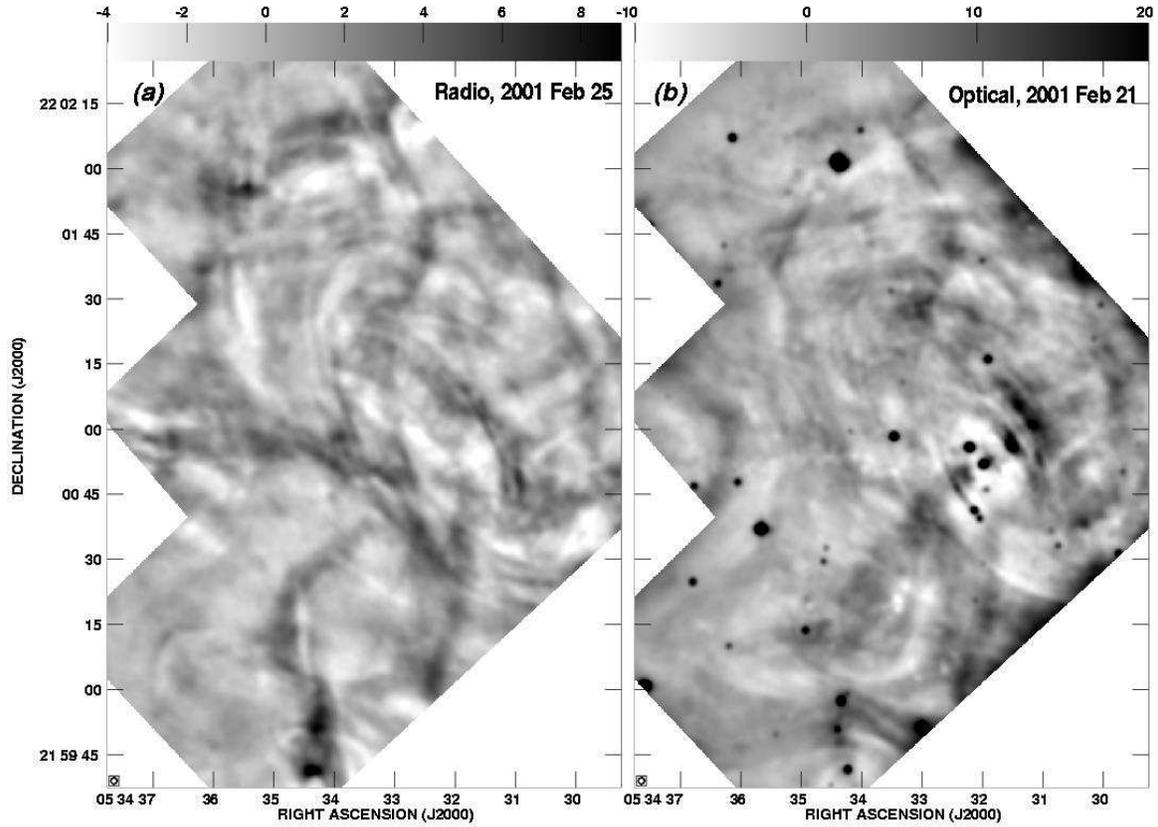}

%  This is 'O2 ALIGN L1.HGEHPB.1', (descended from plane 1 of 
%           O2 ALIGN .OHGEO.2); pixr -10,20 counts
%  and 'CRB-ALLM L4 .CUBBHB.   1' which is only layer 4 of the 7layer
%C     radio cube (only 1 layer was blanked for this purpose)
%C   (0.15" /pixel; 1300x1500pix images); pixr -4,9 mJb
%C    (check/rename 03dec4)
\caption{($a$) The radio image from 2001 February 25 compared with
($b$) the optical image from 2001 February 21, both convolved to the
same effective resolution of 1\farcs4 (FWHM; shown in lower left in
each panel).  Both images have been high-pass filtered with a
14\arcsec\ FWHM Gaussian filter.  The radio image has been blanked to
correspond to the area covered by the \HST\ CCDs on the optical image.
The radio greyscale is labelled in m\Jb, and the optical one in
counts.  Some edge effects due to the high-pass filtering are visible
on the optical image.
% *** This corresponds to video3.mpg ***
%C corresponds to crab-radio-optical.mpg = r+o2.mpg
The accompanying animation shows a region of 61\arcsec\ $\times$
70\arcsec\ near the center of both the radio (left panel) and optical
(right panel) images between 2001 February and April (see
Table~\ref{tobs}). The time-scale is linear.
\label{fr-ohp}}
\end{figure}

\begin{figure}
\epsscale{1.}
%C \plotone{crab-ro-col4a.eps}
%C  crab-ro-col4a seemed the best on our laser printer.....
%C
\plotone{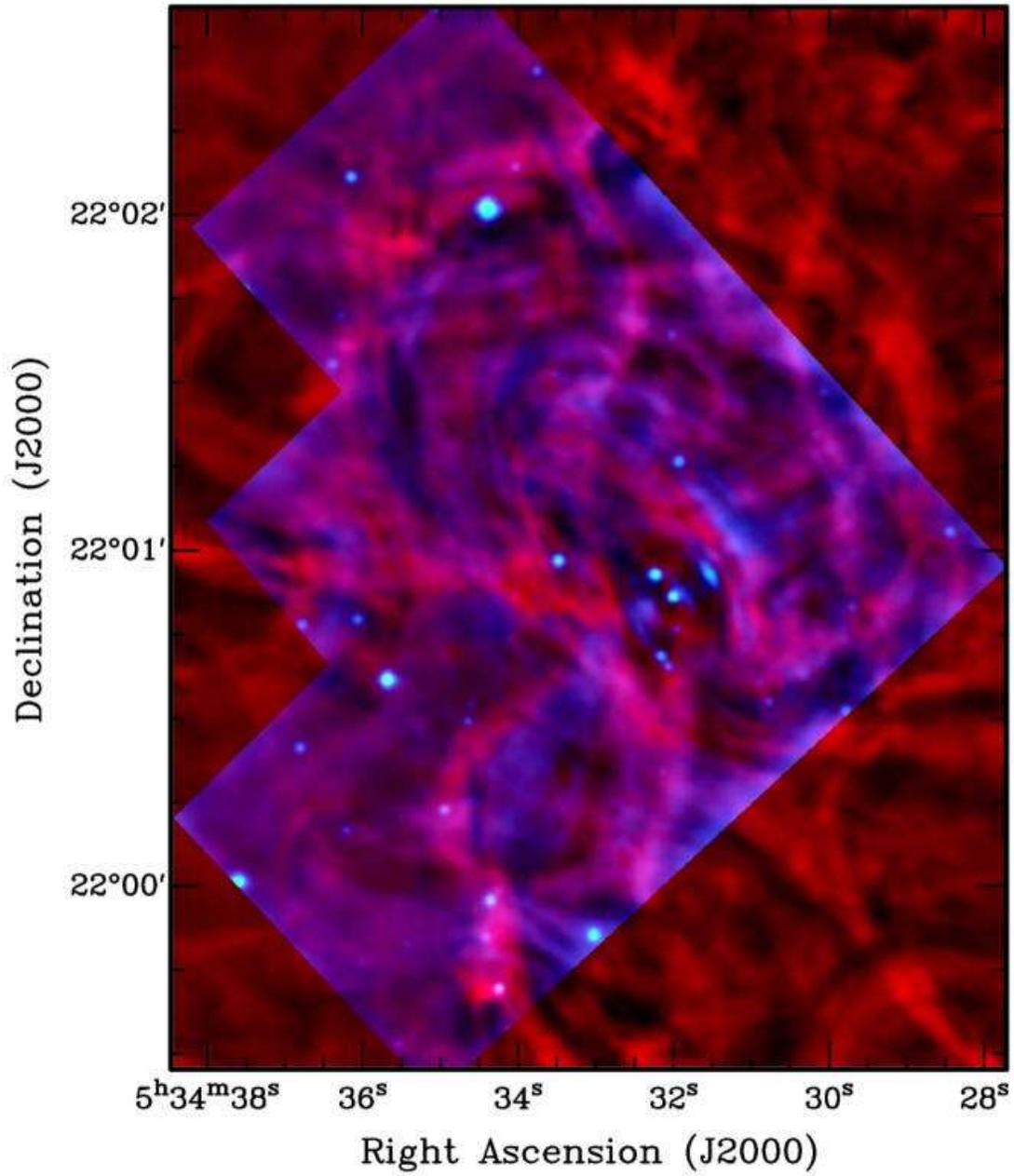}
%C  this last is the one that looked best on the inkjet printer (crab-ro-col4c)
%C  subsequently slightly tweaked to make bluer and 1pixel blur
%
\caption{An overlay of the radio image of 2001 March 27 and the optical one
of 2001 March 26,
convolved to the same effective resolution of 1\farcs4.  Both
images have been high-pass filtered with a 14\arcsec\ FWHM Gaussian
filter.  The radio image is in red and the optical image is in blue
and green.  Some edge effects due to the high-pass filtering are
visible on the optical image.
\label{fr-o-col}}
\end{figure}

We show first an almost simultaneous pair of images of the whole
region covered by the \HST\ CCDs, which is smaller than the radio
images.  Figure~\ref{fr-ohp}$a$ shows high-pass filtered radio image
from 2001 February 25, and Figure~\ref{fr-ohp}$b$ shows the optical image
from 2001 February 21, convolved to the same effective resolution and
similarly high-pass filtered.  The animation accompanying
Figure~\ref{fr-ohp} shows a side-by-side comparison of the central part of
the radio and optical images between 2001 February and 2001 April.
We also show false color overlays of the radio image of 2001 April 16
with the optical images of 2001 April 17 in Figure~\ref{fr-o-col}.

In the outer part of the nebula there is generally a good
correspondence between the radio and the optical, with radio
counterparts existing to most optical features, with the obvious
exception of the stars in the field.  Most of the smaller scale
structure seems to be common to the radio and the optical.  The radio
image contains some additional filaments, not visible in the optical,
which are presumably in front of, or behind, the optically bright
region, since the radio nebula is larger than the optical.
%C remind reader that radio neb is bigger than optical
%C and what importance of correspondence is needed

The correlation between the radio and optical brightness is not as
good near the pulsar as in the outer part of the nebula.  To
illustrate this in a statistical fashion we determined the correlation
coefficient between the optical and the radio brightness.  We sampled
our images every 0.9\arcsec\ in both R.A. and decl.\ for this
comparison, excluding regions near the edge of the \HST\ CCDs and
around the bright stars, and then determined the correlation
coefficient between radio and optical brightness in various regions.
Outside the wisp region, the correlation coefficient between the radio
and optical brightness is 0.41 ($n_{\rm sample} \sim 30,000$) while
inside the wisp region it is 0.30 ($n_{\rm sample} \sim 3000$).
%C  real numbers were 33737 and 2738
%C  wisp region is blc 559 589; trc 763 829 on the 1300x1500 0.15"/pix
%C  aligned images; above stats are for sampling evey 6th pix.

We first compare the general nature of the variability in the wisp
region in the radio with that in the optical by examining the rms over
our images between 2001 February and April.  In Figure~\ref{fro-rms}
we show images of the rms variation with time ($a$) in the radio (four
images) and ($b$) in the optical (six images).  The pulsar is visible in the
radio rms image as a faint point source because, as mentioned, its
average radio flux density is variable.  The strongest radio
%C  say average over time?
variability occurs $\sim$20\arcsec\ to the southwest (SW) of the
pulsar, but radio variability is seen throughout the central region.
The strongest optical variation is that of the bright wisps
$\lesssim$10\arcsec\ to the northwest (NW) of the pulsar, and the
optical variability is concentrated also in the region to the NW of
the pulsar although the base of the optical jet is also visible as a
region of enhanced variability to the SE of the pulsar.
%C average radio brightness over the region shown in
%C Figure~\ref{fro-rms}$a$, before high-pass filtering, was 29~m\Jb
The peak in the radio rms image represents a fractional variability of
$\sim$3\% in the brightness integrated along the line of sight, while
that in the optical represents a fractional variability of $\sim$7\%.

\subsection{Radio Optical Comparisons near the Pulsar}
\label{srocenter}

We will search for a detailed correspondence between the radio and the
optical wisps, and seek to answer the important question of whether
the radio wisps are at the same locations as the optical ones, and
whether they move with the same speed.

%C
%C c = 0.0866 "/day at 2kpc
%C
%C its actually 0.076c pm 0.02 -> 0.175c
%C              0.056c         -> 0.124c
%C              0.096c         -> 0.230c --> 0.18 pm 0.05 c

\begin{figure}
\plotone{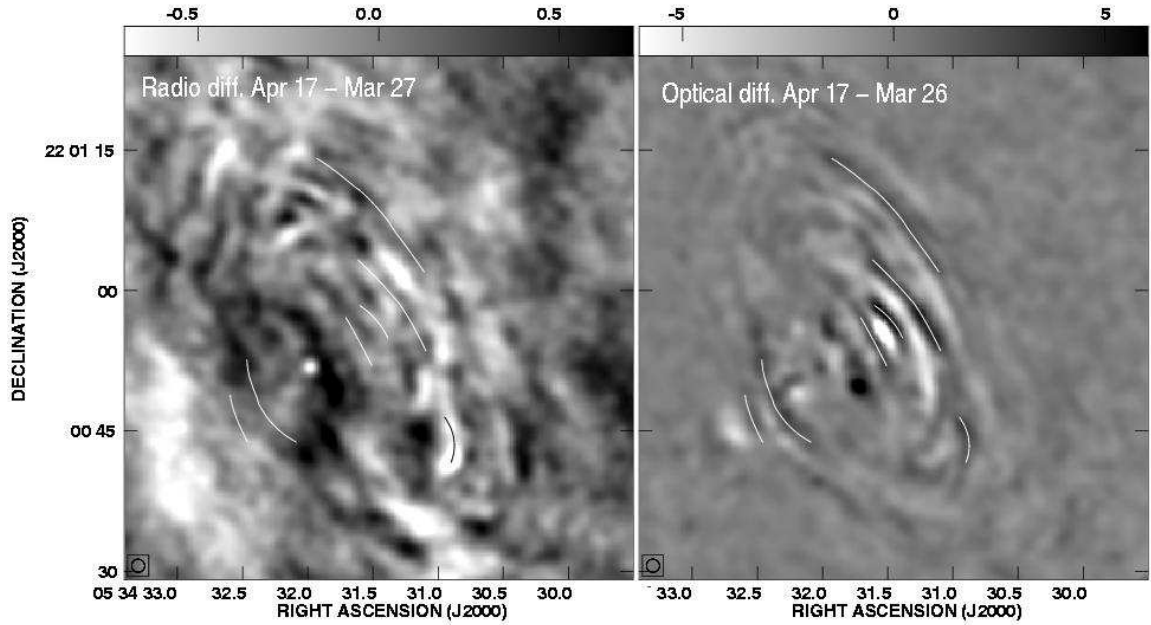} 
%  CRB-ALLM L45 .LYRDFC  etc and O2 L13 .LYRDFC. etc
%C  (0.15"/pix) blc 530,549; trc 890,920
%C  optical,radio added up by radio-opt-diff1.fig etc
\caption{The differences between pairs of consecutive radio images
on the left, with the nearest corresponding optical difference images
on the right.  Both the radio and optical images have been convolved
to an effective resolution of 1\farcs4 FWHM (shown in the lower left
in each panel) and high-pass filtered with a 14\arcsec\ FWHM Gaussian
filter.  The radio greyscales are labelled in m\Jb.  The wisps are
more well defined in the optical, and accordingly we mark the most
prominent positive difference ridges in the optical image with white
lines.  The corresponding lines are drawn in the radio image to show
the location of the optical features in the radio images, switching
from white to black lines where necessary for higher contrast,
\label{fdiffs}}
%C corresponds to craboc-ag-diffbox.ps
\end{figure}

To highlight the rapidly moving wisps, we present further difference
images, this time between pairs of successive images of the closely
spaced observations in 2001.  We form both radio and optical
difference images, matching the times over which the differences are
formed as closely as possible.  We show pairs of radio and optical
difference images in Figure~\ref{fdiffs}.  On such difference images,
motion of a narrow feature results in a pair of parallel positive and
negative ridges, with the positive one on the side toward which the
motion occurs.  In order to compare the precise location of radio and
optical features, we mark the prominent positive optical difference
ridges, i.e., locations where the optical emission has brightened, in
both the radio and the optical images.  We indicate the optical
difference ridges in preference to the radio ones because they are
generally more sharply defined.

The top panel shows the difference between the 2001 March 13 and
February 25 radio images, compared with the difference between the
March 15 and February 21 optical images.  We mark four outwardly
moving optical wisps to the NW of the pulsar and one to the S\@.  The
outermost wisp in the NW has a faint radio counterpart, at least
within our resolution of 1\farcs4.  The next optical wisp seems
distinctly displaced from a radio wisp with a similar geometry, with
an optical positive ridge lying $\sim$1\farcs4 outside the radio one.
%C just measured off of the relevant figure
The remaining two optical wisps, which are the brightest ones, have no
clear radio counterparts.  The optical wisp to the S does seem to have
a faint radio counterpart.  There are also prominent radio features
without obvious optical counterparts, particularly to the SW
of the pulsar.

The middle panel shows the difference between the 2001 March 27 and
March 13 radio images, compared with that between the March 26 and
March 15 optical images.  The pulsar is clearly visible as a positive
point source in the radio image but not in the optical (see
\S~\ref{srocomp}).  As in the previous set of difference images, the
bright optical features nearest the pulsar do not have obvious radio
counterparts.  The outer, northern wisps seem to exhibit some
displacement between the radio and the optical, including some
anti-correlation where positive radio ridges correspond to negative
optical ones. There is also a prominent radio ridge to the
SSW of the pulsar which does not have an optical
counterpart.
%C preceeding might want a bit of work

Finally, the bottom panel shows the difference between the 2001 April
17 and March 27 radio images, compared with that between the April 17
and March 26 optical images. The pulsar has become fainter in the
radio, and is visible as a negative source in the radio image.  The
marked wisp to the SW of the pulsar shows a clear anti-correlation,
being negative in the radio and positive in the optical, that is
having become fainter between April 17 and March 27 in the radio, but
having become brighter over almost the same interval in the optical.
As in the previous set of difference images, the bright optical
features near the pulsar do not have obvious radio counterparts, and
there are also radio features without optical counterparts.
%C -- I seem to have written this at some point but its wrong, just say nothing
%C The second wisp from the outside to the NW also
%C seem to show an anti-correlation, with a corresponding radio features
%C with the opposite sign.  In this case it is not clear whether the
%C radio feature is outside or inside the corresponding optical feature.

\subsection{Is there a Radio Counter-Part to the Optical Jet}
\label{sjet}

A jet is visible in both the optical and the X-ray images, and movement
along it can be seen at speeds of at $0.3 \sim 0.4c$ (Hester et~al.\
2002; Mori et~al.\ 2002).  The geometry of the jet in the optical is
similar to that in the X-ray: the base of the jet is bright, it extends
approximately 1\arcmin\ to the SE of the pulsar, becoming wider and
curving slightly to the S toward its end.  The base of the jet is visible
as a region of variability $\sim$10\arcsec\ to the SE of the pulsar in
the optical rms image (Figure~\ref{fro-rms}$b$).  There does not seem
to be an obvious radio counterpart to the bright base of the jet,
either in the radio rms image (Figure~\ref{fro-rms}$a$) or in the
images of individual epochs (e.g., Fig.~\ref{fr-ohp}$a$).  A highly
variable feature called the ``sprite'' is seen at the base of the jet
in the optical (Hester et al.\ 1998).  Although there is a weak,
diffuse radio feature roughly at the location of the sprite, the
positional alignment is not exact and the radio feature is not
especially variable, so it cannot reliably be identified with the
sprite.

%C  optical Jet (hi-pass, convolved) rms ~2, avg. 75 counts, 3%
%C   radio jet rms ~ 200 uJy, avg. 30 mJy/bm -> 0.6%
%C  so optical fractional variability of jet is 5x higher
%C   if we take out overall factor of 2 because of line-of-sight,
%C   its a little more than 2, not v. conclusive.
%C  Also the radio jet does line up rather nicely with the end of 
%C   the optical jet.  Need to take a better look at X-ray.

%C  Dale says remove whole radio jet

\section{Discussion}
\label{sdiscuss}

We have produced time-resolved high-resolution radio images of the
Crab Nebula which are almost simultaneous with similar optical and
X-ray images.  We confirm the existence of the rapidly moving radio
wisps in the central 2\arcmin\ of Crab Nebula, first detected as
spectral index variations by Bietenholz \& Kronberg (1992), and
subsequently identified as mobile features in Paper~I\@.  We show that
the rapid motions of the radio wisps are definitely directed outward,
with projected speeds of $\lesssim 0.3\, c$.  In addition to the wisp
motion, there is complex radio variability over longer time scales of
$\sim$3~year, including features farther from the pulsar which move
much more slowly than the wisps, but still more rapidly than the
nebular expansion. We compared the morphology and motions visible in
the radio to those visible in the optical synchrotron emission.  We
now discuss the relevance of these motions in different regions of the
nebula in more detail.

\subsection{The Wisps and Particle Acceleration in the Crab Nebula}

What insight can the observed behavior of the radio and optical wisps
give into the establishment of the nebular particle spectrum and in
particular the question of the origin of the radio-emitting electrons?
We proposed in Paper~I that the rapidly variable radio wisps are
associated with the optical wisps and the X-ray torus.  The fact that
the radio and optical wisps show a similar overall geometry of
elliptical arcs, and the fact that both are seen to move away from the
pulsar with speeds which are substantial fractions of $c$ suggests
that the radio wisps are intimately associated with the optical ones.
The elliptical geometry suggests that the wisps are intrinsically
circular features in or near the equatorial plane of the pulsar.  The
fractional time variability in the radio brightness is of the same
order of magnitude as that of the optical brightness, suggesting that
the radio-optical spectrum of the time-variable component, i.e., the
wisps, is roughly comparable to that of the bulk of the
nebula\footnote{The fractional variability in the radio brightness is
approximately half of that in the optical.  However, the radio
brightness along lines of sight through the center of the nebula
contains considerable emission from regions in front of and behind
the optically bright region, since the radio synchrotron nebula is
larger than the optical one.  Since these optically faint regions far
from the pulsar are not likely to be highly variable, it seems
probable that the fractional time-variability in the radio would be
similar to that in the optical, i.e., on the order of 1\%, if one
considers similar emitting volumes.}. This further implies that the
particle spectrum in the wisps is also similar to that in the bulk of
the nebula, at least over the range of energies of the
particles responsible for the radio and the optical synchrotron
emission.

While this evidence of association between the radio and the optical
wisps is circumstantial, we feel that it is compelling.  The
simultaneous observations, however, show that the correspondence
between radio and optical wisps is not exact.  While some wisps are
bright in both radio and optical, others are bright in the radio and
faint in the optical or vice versa.  This lack of detailed
correspondence argues for two different but related processes
producing the radio and the optical wisps. By contrast, the
correspondence between the optical and the X-ray wisps is good (Hester
et al.\ 2002), suggesting that the X-ray wisps are produced by the
same process as produces the optical ones.
%Cxxx Do H et al, what about Mori? say anything about this.

A long-standing problem, mentioned in the introduction, is the
provenance of the radio emitting electrons, i.e., those with Lorentz
factors of $\lesssim 10^4$. 
%C The visible nebula is powered by the spindown energy
%C of pulsar, which emerges in the form of a collimated, highly
%C relativistic wind consisting of some combination of particles and
%C magnetic field.  
In most theoretical pictures of the Crab, the bulk Lorentz factor of
the pulsar wind, $\gamma_{\rm w}$, is expected to be $\sim10^6$
(e.g., Wilson \& Rees 1978; Kennel \& Coroniti 1984a,~b; Melatos \&
Melrose 1996; Arons 2002).
%C Wilson and Rees claim (according to KC1984) that lower limit is
%C 10^4 because otherwise forced compton scattering would smear out
%C radio pulses.  Arons (private comm) says presence of B modifies
%C this but KC1984 don't say whether up or down!  Arons himself
%C never cites the Wilson & Rees paper according to ADS, so possibly
%C he thinks its total hooey.
The wisps are associated with a shock in the equatorial region of the
pulsar wind.  It is in, or immediately downstream from, this shock
that the nebular particle spectrum is established.
%C
The problem posed by the radio-emitting electrons is that it is
difficult to produce a substantial number of electrons with $\gamma$
much lower than $\gamma_{\rm w}$ in such a shock.  Thus, if
$\gamma_{\rm w}$ is in fact $\sim 10^6$, then the numerically dominant
radio-emitting electrons with $\gamma \lesssim 10^4$ are difficult to
account for.  Furthermore, if $\gamma_{\rm w} \sim 10^6$, then the
total number of relativistic electrons currently injected is far too
small to account for the number of radio-emitting electrons currently
observed in the nebula, even allowing for the fact that the synchrotron
lifetimes of the lower-energy electrons exceed the age of the nebula.

One possible explanation for the existence of the radio-emitting
electrons is that they are of historical origin (e.g., Atoyan 1999).
%C was suggested but probably dismissed in Kennel & Coroniti 1984b
Since the synchrotron lifetimes of the $\gamma \lesssim 10^4$
electrons are longer than the age of the nebula, they could have been
injected at some point in the past.  A second possibility is that
there may be a second outflow process which is responsible for
producing the low energy electrons --- perhaps the bulk of the energy
emerges in an equatorial wind with $\gamma_{\rm w} \sim 10^6$, but
there is an additional, slower, outflow, perhaps at higher latitudes,
which is the source of the radio-emitting electrons. In neither of
these cases, however, would radio emission currently be expected from
the wisp region.  Since we do observe radio emission from the wisp
region, neither of these two scenarios seem likely to
apply.\footnote{We note that a population of electrons with a
low-energy cutoff such that $\gamma > \gamma_{\rm min}$ will produce
synchrotron emission with a low frequency tail whose brightness is
$\propto \nu^{1/3}$.  However, the radio emission from the wisps is
far too bright to be merely the low frequency tail for any reasonable
value of $\gamma_{\rm min}$, for example $\gamma \sim 10^5$.}
%C dale says can we get a ref, not that important

The association of the radio and the optical wisps suggests rather
that there are two closely related processes which produce the radio
and the optical wisps, and also the low- and the high-energy
relativistic electrons.  Two candidates for a two-part process of
particle acceleration have been put forward.  In both cases, the
high-energy particles are produced by Fermi acceleration, which seems
capable of producing the required particle spectrum.  Gallant et~al.\
(2002) hypothesized that the pulsar wind contains ions in addition to
electrons, and that the resonant absorption of ion cyclotron waves
accelerates lower energy electrons, as was first suggested by Hoshino
et~al.\ (1992).
%C It is unclear whether the required number of radio emitting electrons
%C could be produced in this manner.
Recently, Lyubarsky (2003; see also Lyubarsky \& Kirk 2001), showed
that the low energy part of the particle population can be produced by
rapid, forced, reconnection of the alternating magnetic field stripes
in the termination shock of the pulsar wind.  In this case, the pulsar
wind before the termination shock is Poynting dominated, but a
significant fraction of the energy is transferred to the particles
immediately after the shock because of reconnection, with the result
that the average particle energy grows as the particles pass through
the shock.  This would imply a bulk Lorentz factor in the wind of only
$\sim 4.4\ex{3}$ for an injection rate of $10^{40}$~\sn{-1}.  Some
separation between the radio and optically bright wisps could be
accommodated in either of these scenarios, but neither makes particular
predictions as to the relative locations of the radio and optical
wisps.
%C Assuming a similar spectrum to the bulk of the Nebula ($\alpha = -0.3$
%C between 100~MHz and 100~GHz), this represents a radio luminosity of
%C $\sim 10^{33}$~erg~\sn{-1}.
%C  4 pi r**2 * int[(nu/5G)**-0.3]100MHz->100GHz =
%C  4 pi r**2 *  nu**0.7/0.7|1e8->1e11 (5e9)**0.3  
%C  4.818e44(cm*2) * 5.769e10 
%C  2.780e55 (f in cgs) = 2.780e32 (f in Jy)
%C    whole radio lum. is 2.2e35 -> wisps 4.4e33
%C  check number for radio lum

The physical nature of the wisps is also a matter of some debate
(e.g., Hester et al 1995; Chedia et al.\ 1997; Arons 1998; Lou 1998;
Begelman 1999; Komissarov \& Lyubarsky 2003; Spitkovsky \& Arons
2004).  Unfortunately none of the models to our knowledge directly
address the issue of the lower-energy electrons, or make direct
predictions concerning radio wisps, largely because of the
aforementioned problem of the origin of the lower-energy electrons.
We do not discuss the various theories for the origin of the wisps in
detail, but list some observational constraints provided by the
observed differences between the radio and optical wisps.  The optical
wisps to the NW of the pulsar are considerably brighter and move more
rapidly than those to the SE, which is plausibly explained by Doppler
beaming of material moving out from the pulsar with at least
transrelativistic velocities, with the bright parts of the wisps to
the NW being those where the motion is toward us.  In the case of the
radio wisps, by contrast, those to the NW of the pulsar are
not notably brighter than those to the SE.  The observed projected
velocities for the radio wisps are also somewhat lower than those of
the optical wisps.  This suggests that the bulk flow in the radio
wisps is slower.  The somewhat more amorphous nature of the radio
wisps may further suggest that they are somewhat more widely
distributed above and below the pulsar's equatorial plane than the
optical ones.  In particular, the brightest optical wisps near the
pulsar do not seem to have radio counterparts.  Perhaps, the
establishment of the low energy particle population is slower than
that of the higher energy one, and thus the number of radio emitting
electrons is still low in the wisps closest to the pulsar.
%C  Optical Larmor ~ 10^14 cm = 3 mas;  
%C  Radio   Larmon ~ 10^9  cm = v. tiny
%C or be due to a relatively fine segretation in the flow
%C 

%C  Larmor radius
%C  Astroph formula: Larmor radius = m c gamma v_perp / q B
%C  assume that Gauss, esu works ->  
%C   r = (9.109e-28 g)(3e10 cm/s)^2 (gamma)(v_perp/c) /
%C             (4.802e-10 esu) 1e-3G (B in mG)
%C     = 1.71e6 cm * gamma * (v_perp/c) / (B in mG)
%C    so for b = 0.4 mG, gamma = 300, v_perp/c = 0.3 ->
%C    r = 3.84e8 cm
%C  Lyubarsky gets only 10^12 cm
%C  Arons get 0.15pc for ions -> 4.7e17cm, so 2.5e14 for e-
%C 

\subsection{The Moving Arc}
\label{sarc}

The feature we have termed the moving arc, described in
\S~\ref{sradvar} above, is prominent in the $\sim$3~year difference
image (Figure~\ref{fdifimg}$a$).  It had already been seen in the
radio, showing distinct changes in brightness between two sets of
1.4~GHz observations in 1982 and 1987 (Bietenholz \& Kronberg 1992).
We note that the speed of its outward motion we estimated is higher
than that estimated by Bietenholz \& Kronberg, which would imply
substantial acceleration of the moving arc.  However, we measured the
speed of the moving arc from the part with the largest changes in
brightness, while Bietenholz \& Kronberg measured the speed of the
edge toward the pulsar, so a direct comparison of the speeds is
probably not warranted.
The moving arc is also apparent in the optical: it is clear in an
optical difference image, between observations in 1988 and 1994
(Hester et al.\ 1995), with a very similar morphology as is seen in
the radio.  It probably also corresponds to the ``dark lane''
described by Oort \& Walraven (1956) and Scargle (1969), and possibly
already observed by Lampland (1921).  Its present location and speed
is approximately consistent with that determined by Scargle (1969),
suggesting a secularly evolving feature moving outward through the
nebula at a projected speed of order $10^4$~\kms, about ten
times that of the nebular expansion.

The moving arc is approximately opposite the pulsar from the jet seen
in the optical and X-ray.  Might it be associated with a counter-jet?
Although a counter-jet is not clearly seen at any wavelength, there is
emission identified as the counter-jet channel in both the optical and
the X-ray (Hester et al.\ 2002).  The uniqueness and prominence of the
moving arc suggest a possible association with the distant end of the
counter-jet.  The alignment of the moving arc with the counter-jet is
not exact, but since the jet is seen to be curved, it seems likely
that the counter-jet is curved also.
%C If the moving lane is associated with the counter-jet, we can
%C conclude that the bulk flow is no longer relativistic, since otherwise
%C the counter-jet would be faint due to Doppler beaming.  
%C The observed proper motion of the arc, which is partly transverse,
%C might be due to instability in the jet flow.

A serious problem with the association of the moving-arc with the
counterjet is the absence of an at least similarly prominent radio
feature associated with the jet.  There is in fact a less prominent
feature visible approximately opposite the moving arc in
Figure~\ref{fdifimg}$a$, however its association with the jet is
entirely speculative at this point.  Possibly the prominence of the
moving arc in the NW over any feature in the SE is due to the more
rapid present motion of the counter-jet, since the moving arc is
prominent only in difference images.  However, any association between
radio features and the optical and X-ray jet and its implied
counter-jet must remain speculative until future observations show a
clear connection.

\subsection{Other Motions in the Outer Nebula}
\label{soutmotion}

On the $\sim$3~year difference image in Figure~\ref{fdifimg}$a$,
striations are visible in most of the nebula, which suggest wave-like
motions.  They are not clearly discernible enough to determine the
speeds, but proper motions in the range of $0.2 \sim
1$\arcsec~year$^{-1}$ are suggested, implying projected speeds of $2
\sim 10 \times 10^3$~\kms.
%C  1"/yr = 2.992e16cm/yr = 9481 km/s (2.0 Kpc)
%C  or c = 0.0834"/yr
Similar motions are reported in the optical in a $\sim$6~year
difference image (Hester 1995).  The morphology suggests traveling
waves, perhaps reflecting off the boundary of the nebula and
diffracting from the massive filaments of thermal gas.  Since the
wisps and the jet, where the synchrotron fluid is injected into the
nebula, are seen to be unsteady on short time-scales, it seems
reasonable to suppose that wave-like disturbances, excited by the wisp
and jet instabilities, would propagate away from the injection region
in the synchrotron-emitting plasma, where the sound and Alfv\'en
speeds are relatively high.

\section{Summary}

By studying time-resolved radio observations of the Crab Nebula, and
comparing them to simultaneous and similarly time-resolved optical and
X-ray images we have found the following:

\begin{trivlist}
\item{1.} We confirm the existence of radio-emitting counterparts to
the optical wisps. We show that the radio wisps, like the optical
ones, have arc-like geometries and are moving outward at speeds
of up to $\sim0.3c$.

\item{2.} The fractional variability of the brightness at the wisp
locations in the radio is similar to that in the optical, implying
that the particle spectrum in the wisps is similar to that in the body
of the nebula.  This suggests that the radio-emitting electrons with
$\gamma < 10^4$ are accelerated in the wisp region along with those
with higher energies.

\item{3.}  The correspondence between the radio and the optical wisps
is not exact.  In several cases the radio brightness seems to be
anti-correlated with that in the optical.  In particular, the
brightest optical wisps nearest the pulsar do not seem to have bright
radio counterparts.

%C Q are the fast motions to the NW of the psr?  make sure were
%C not confusing variable and stationary features

\item{4.} The jet seen in the optical and X-ray does not have an
obvious radio counterpart.

\item{5.} The lack of detailed radio-optical correspondence suggests
two different but related processes for producing the radio and
the optical wisps, and the populations of relativistic electrons
with $\gamma$ below and above $\sim 10^4$.

\item{6.} There is a prominent, curved, moving radio feature in the NW
of the pulsar that we call the moving arc.  It's projected
speed is on the order $10^4$~\kms, at least an order of magnitude
higher than those entailed by the general expansion of the nebula.
This is a long-lived feature: it was detected almost 20 years ago in
the radio, and at least 30 years ago in the optical.

\item{7.} There is considerable variation of the radio emission with
time-scales of $\gtrsim$1~year, which is longer than those associated
with the wisps.  On these time-scales, the fractional variability of
the radio brightness near the pulsar is $\sim$1\%.

\item{8.}  Striation is seen in the radio difference images spanning
$\sim$3~year, suggesting the presence of wave-like motions in the body
of the nebula, albeit with lower amplitudes than the wisps and the
moving arc.  The geometry is generally arc-like but not centered on
the pulsar like that of the wisps.

\end{trivlist}

\medskip
\acknowledgements Research at York University was partly supported by
NSERC\@.  We thank Barry Clark for his continued patience with the
scheduling difficulties involved in these observations.  R.
Bartel assisted with the data reduction.
  
%  all refs in text & vice versa checked July 13, 2004

\clearpage

%C  Color figure generation:
%C  Lets use 26Mar for this one - we have both on same
%C       day.  Optical=layer 4; radio = layer6
%C   #4: kview-rgb to overlay crab-allm-lyr6.fits (red)
%C       and o2al-hgehpc-lyr4b.fits  (both blue and green)
%C       in kview set blue pixr to about lower half, green to upper
%C       set label-color to black (dissappears on screen) for eps
%C       then read into gimp at 300dpi, weak anti-aliasing;
%C       crop- autoshrink it
%C       tweak contrast a little
%C       and use rotate-colormap to expand red-cyan color range 
%C       4a from 0-1 to 0.5 to 0.09
%C       4b from 0-1 to 0.33 to 1 reverse order
%C   #5  as above but blue-green is o2al-hgehp-lyr4b.fits
%C       (optical unconvolved; DATAMAX changed to 75); only 5a made.
%

%C  alternate colour figures
%C \plotone{crab-ro-col4gimp.eps}
%C \plotone{gimp-rgb3.eps}
%C \plotone{crab-ro-col5a.eps}

\begin{figure}
\plotone{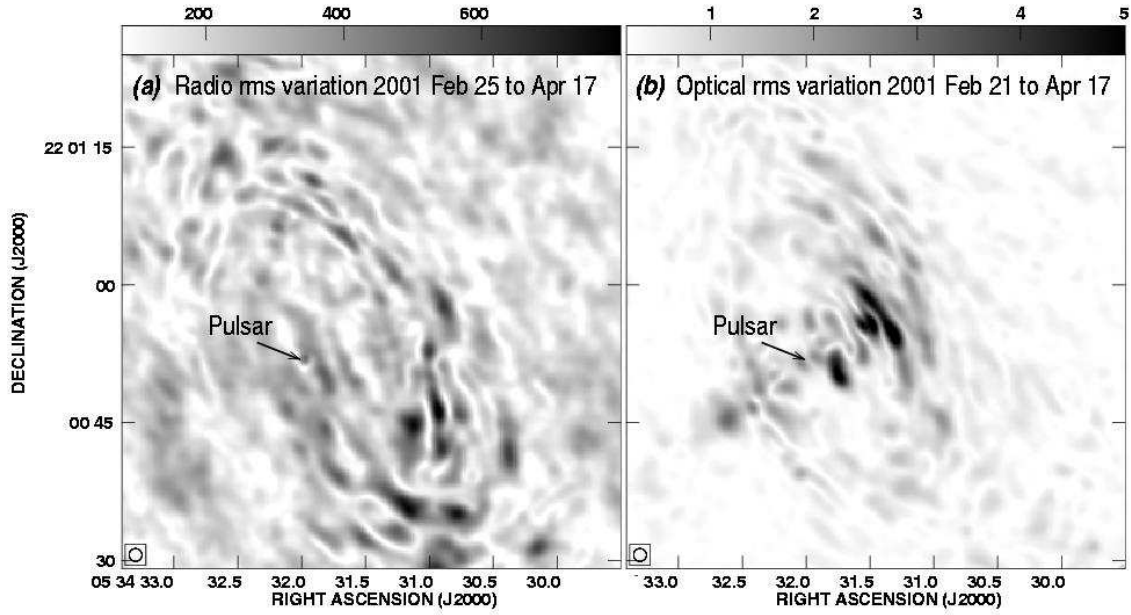} 
\caption{The rms variation with time over the 2001 February to 2001
April observing period.  Both panels show the rms variation with time
of images convolved to 1\farcs4\ (indicated at lower left) and
high-pass filtered with a Gaussian of FWHM 14\arcsec.  The pulsar
position is indicated on both panels.  ($a$) The rms variation over
the four high-pass filtered 2001 radio images, with the greyscale
labelled in $\mu$\Jb.  ($b$) The rms variation over the optical images
of 2001 February 21, March 04, 15, 26, April 6 and 17, with the
greyscale labelled in counts.
%C (see Hester et al.\ 2002) 
\label{fro-rms}}
\end{figure}

\begin{figure*}
\epsscale{1.0}
\plotone{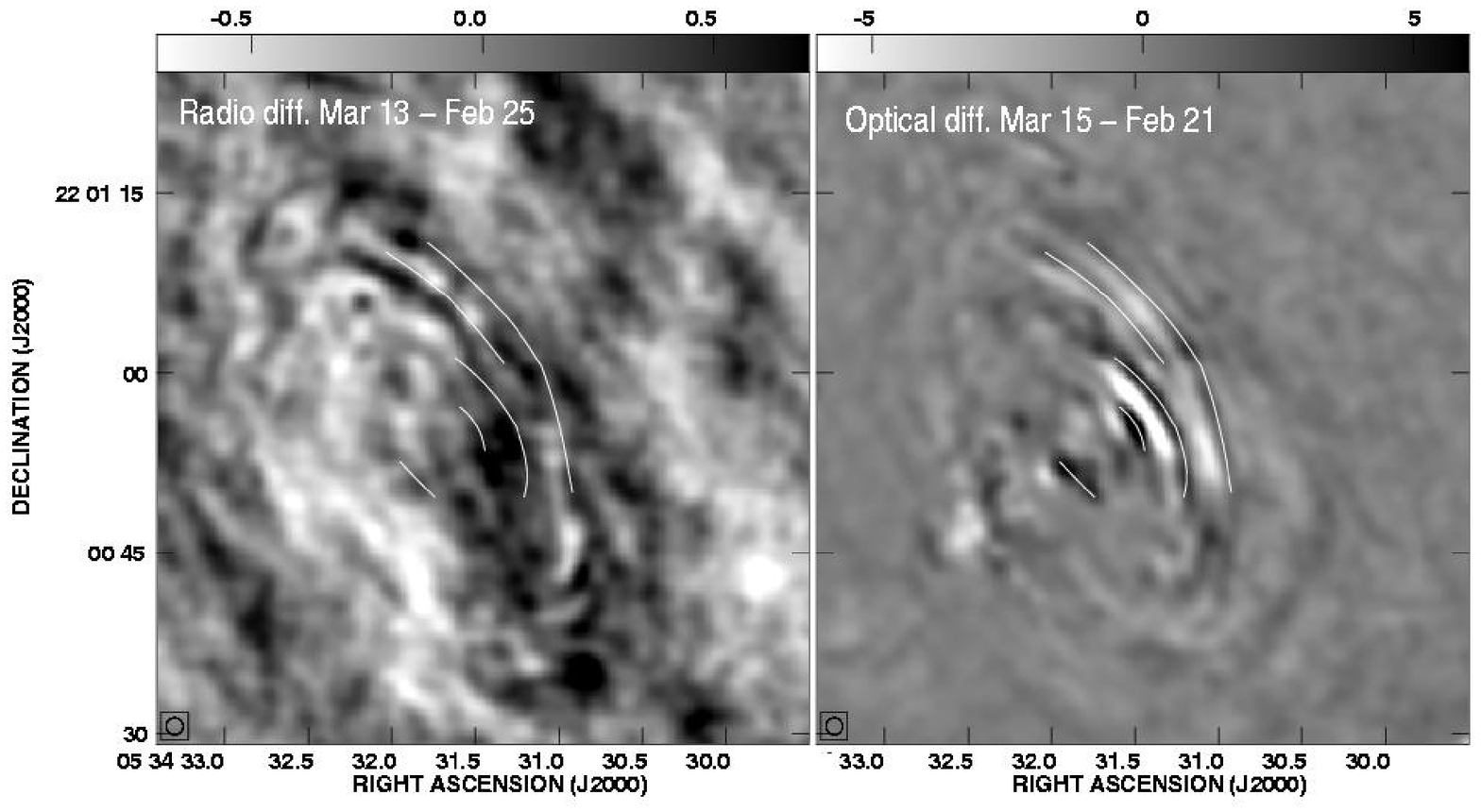} 
\plotone{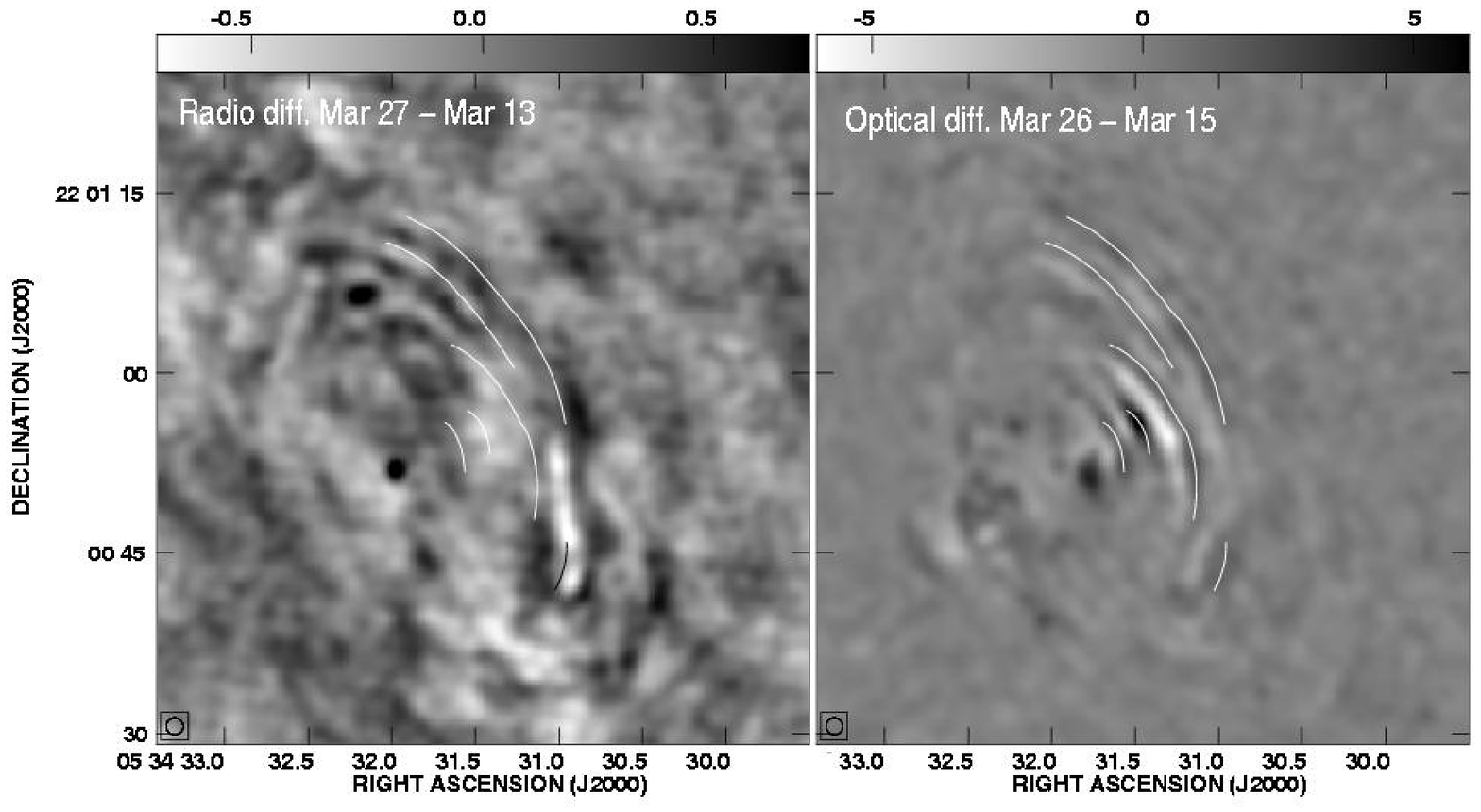}
\end{figure*}

\end{document}